\documentclass[
 reprint,
 amsmath,amssymb,
 aps,
floatfix,
]{revtex4-2}
\usepackage{graphicx}
\usepackage{dcolumn}
\usepackage{bm}

\usepackage{hyperref}

\begin{document}

\preprint{APS/123-QED}

\title{Hierarchical Community Detection in Bipartite Networks:}
\thanks{A footnote to the article title}%

\author{Tania Ghosh}
\affiliation{Department of Physics, University of Houston, Houston, TX 77204.}
\affiliation{Texas Center for Superconductivity, University of Houston, Houston, TX, 77204}
\author{Kevin E. Bassler}%
 \email{bassler@uh.edu}
\affiliation{Department of Physics, University of Houston, Houston, TX 77204}
\affiliation{Texas Center for Superconductivity, University of Houston, Houston, TX, 77204}
\affiliation{Department of Mathematics, University of Houston, Houston, TX 77204.}



\date{\today}

\begin{abstract}
Many bipartite networks exhibit hierarchical community structure, but existing community detection methods are not well-suited for detecting hierarchy. They also do not effectively handle weighted bipartite networks.
In this work, we introduce a novel modularity-based objective function, called the generalized bipartite modularity density, $Q_{bg}$, specifically designed for hierarchical community detection in bipartite systems. The framework incorporates a tunable resolution parameter that enables systematic exploration of community structure across multiple scales. It leverages resolution-limit behavior in bipartite networks as a tool to uncover hierarchical organization without projecting the network or altering its intrinsic bipartite topology.
We evaluate the method using a hierarchical synthetic bipartite benchmark and apply it to two empirical networks. In all cases, $Q_{bg}$ recovers established mesoscale structure while revealing additional hierarchical and fine-scale organization beyond that detected by conventional bipartite approaches. These results establish $Q_{bg}$ as a flexible, interpretable, and resolution-aware framework for hierarchical community detection in bipartite networks.
\end{abstract}


\maketitle


\section{\label{sec:Introduction}Introduction}
Community detection is a central problem in network science, providing insight into the structural and functional organization of complex systems~\cite{Newman_2003,Fortunato2010PhysRep}. In many real-world networks, community organization is inherently hierarchical, with meaningful structure present across multiple scales~\cite{Newman_2003,Reichardt_2006,LU202054}. Bipartite networks form an important class of network representations in which nodes are partitioned into two disjoint sets and edges connect only nodes of different types, often with weights encoding the strength of interactions. Such bipartite structures arise naturally in diverse domains, including social affiliation and collaboration networks~\cite{Barber_2008,Latapy2008SocialNetworks}, social interaction systems~\cite{Newman_2001,Matsuo_2009}, and biomedical data~\cite{Goh_2007,Barabasi2011NetworkMedicine}. In many of these systems, community organization is expected to span multiple scales, motivating the need for methods capable of resolving hierarchical community structure while preserving the intrinsic bipartite topology.

Most existing community detection approaches for bipartite networks are based on modularity optimization. A well-known limitation of such methods is the resolution limit, which can cause small but well-defined communities to be merged into larger clusters~\cite{Bu_2018,Chen_2014,Mehdi_2016}. However, previous work has shown that, rather than being a drawback, the scale associated with the resolution limit can be exploited as a tool to uncover hierarchical structure in unipartite networks by varying a resolution parameter~\cite{Guo_2023}. In that work, a modularity-based objective function, termed the generalized modularity density, was introduced, incorporating a tunable parameter that enables the resolution of community structure at any desired scale.

Building on this idea, we extend the framework to bipartite networks. To this end, we introduce a novel objective function, termed the \emph{Generalized Bipartite Modularity Density}, which incorporates a tunable resolution parameter $\chi$. This parameter enables the systematic detection of communities at different scales, allowing hierarchical community structure to be identified directly within bipartite networks while preserving their intrinsic topology.

The remainder of the paper is organized as follows. In Sec.~\ref{sec:Methods}, we present the proposed framework. We first discuss the resolution-limit problem in bipartite networks, followed by the definition of the generalized bipartite modularity density and its properties. We then analyze resolution-limit behavior using a benchmark network. In Sec .~\ref {sec:Results}, we first evaluate the performance of the method on benchmark tests. We then demonstrate its applicability to both synthetic and real-world bipartite networks, including hierarchical artificial networks, the Southern Women network, and an asthma patient network. Finally, additional analytical details and supporting results are provided in the Appendix.

\section{\label{sec:Methods}Methods}

\subsection{\label{sec:ResolutionLimit_in_Bipartite}Resolution Limit in Bipartite Networks}

In the bipartite setting, the resolution-limit problem was explicitly demonstrated using a synthetic benchmark consisting of a ring of interconnected bipartite cliques~\cite{ABC}. In this construction, standard modularity-based methods incorrectly merge adjacent cliques and fail to recover the planted community organization, illustrating how resolution-limit effects can obscure hierarchical structure even in simple, well-controlled bipartite systems.

To examine this limitation more closely, we analyze the bipartite modularity formulation introduced by Barber~\cite{Barber_2007}, which extends modularity to bipartite networks without requiring one-mode projection. This formulation evaluates the quality of a partition by comparing observed edge weights within communities to those expected under an appropriate bipartite null model. The bipartite modularity introduced by Barber~\cite{Barber_2007} is defined as
\begin{equation}
    Q_b = \frac{1}{F} \sum_{c} \left( m_c - \frac{q_c d_c}{F} \right),
    \label{eqtn:barber}
\end{equation}
where \(F\) denotes the total edge weight of the bipartite graph, \(m_c\) is the total weight
of edges within community \(c\), and \(q_c\) and \(d_c\) represent the total (weighted) degrees
of the two node types within community \(c\).

We consider two competing partitions of a bipartite $k$-clique ring: the \emph{Separated} phase (S), in which each bipartite clique forms an individual community, and the \emph{Merged} phase (M), in which adjacent cliques are grouped into larger modules~\cite{ABC}. Let \(Q_b(S)\) and \(Q_b(M)\) denote the corresponding bipartite modularity values. If
\begin{equation}
Q_b(S) > Q_b(M),
\label{eq:eqtn2}
\end{equation}
then bipartite modularity favors the finer partition; otherwise, it fails to resolve all communities due to the resolution limit.

Our analytic derivation shows that this inequality holds only when
\begin{equation}
k > \sqrt{\frac{n - 2}{2}},
\end{equation}
where \(k\) is the size of each bipartite clique and \(n\) is the number of cliques in the ring. The detailed derivation, which corrects the earlier result reported in~\cite{ABC}, is provided in Appendix~\ref{sec:appendix_resolution}. This demonstrates that Barber’s bipartite modularity is also subject to a resolution limit and fails to recover the planted community structure when \(k\) is small.

\subsection{\label{sec:Gen_BipMod_density}Generalized Bipartite Modularity Density}

We introduce a novel modularity-based objective function, termed the \emph{Generalized Bipartite Modularity Density}, designed for hierarchical community detection in bipartite networks. It is defined as
\begin{equation}
Q_{bg} = \frac{1}{F} \sum_{c} \left( m_c - \frac{q_c d_c}{F} \right) \rho_c^{\chi},
\label{eq:Qbg-definition}
\end{equation}
where \(F\) denotes the total edge weight of the bipartite network, \(m_c\) is the total weight of
edges within co-cluster \(c\), and \(q_c\) and \(d_c\) represent the total weighted degrees of the
two node types within cluster \(c\). The quantity \(\rho_c\) denotes the internal link density of
co-cluster \(c\), defined as
\begin{equation}
\rho_c = \frac{m_c}{n_r n_b},
\label{eq:eq46}
\end{equation}
where \(n_r\) and \(n_b\) are the numbers of nodes of the two types in the cluster. This density
measures the fraction of realized links relative to the maximum possible number of bipartite
connections. The parameter \(\chi\) is a tunable resolution parameter that controls the influence
of internal link density on the modularity score.

By construction, \(Q_{bg}\) reduces to the standard bipartite modularity \(Q_b\) when \(\chi = 0\),
preserving its emphasis on dense intra-community connectivity. For \(\chi > 0\),
the contribution of link density is progressively amplified, enabling explicit control over the
resolution at which communities are detected. This makes \(Q_{bg}\) a natural and interpretable
multi-resolution extension of bipartite modularity.

The formulation in Eq.~\eqref{eq:Qbg-definition} naturally extends to weighted networks by
interpreting \(m_c\), \(q_c\), and \(d_c\) as sums of link weights rather than edge counts, as in
weighted modularity~\cite{Newman_2004,Newman_2008}. However, incorporating link density in weighted
networks requires additional care. A naive approach computes \(\rho_c\) by ignoring weights and
treating the network as unweighted~\cite{Chen_2015}, which discards information about edge strength.
A more consistent definition introduces a weighted density,
\begin{equation}
\rho_c = \frac{m_c}{(n_r n_b) w_{\max}},
\label{eq:eq47}
\end{equation}
where \(w_{\max}\) is the maximum link weight in the network. While this definition preserves
consistency with the unweighted case, it introduces strong sensitivity to extreme weight values.
In particular, when link weights vary widely and the maximum weight \(w_{\max}\) is large,
the value of \(\rho_c\) becomes very small, which can lead to numerical instability in
modularity density--based optimization.

In contrast, the generalized modularity density \(Q_{bg}\) avoids this instability. In
Eq.~\eqref{eq:Qbg-definition}, both terms inside the summation are multiplied by
\(\rho_c^{\chi}\), so the global factor \(w_{\max}^{\chi}\) factors out uniformly across all
partitions. As a result, the optimization of \(Q_{bg}\) is independent of the absolute scale of
edge weights, allowing relative link density to be used without affecting the outcome.

By tuning the parameter \(\chi\), the method flexibly favors either the merging or separation of
communities, depending on the desired level of granularity. As shown below~\ref{ResolutionLimitBenchmark}, for appropriate values
of \(\chi\), maximizing \(Q_{bg}\) can correctly identify distinct communities even in the presence
of substantial interconnectivity. This flexibility makes \(Q_{bg}\) particularly well suited for
networks with hierarchical or multi-scale community structure, which are commonly observed in
real-world systems.

Finally, we evaluate the performance of \(Q_{bg}\) by comparing it with standard bipartite
modularity \(Q_b\) on a challenging benchmark network that extends beyond simple ring-of-cliques
constructions. Our results show that \(Q_{bg}\) consistently outperforms \(Q_b\) across the
benchmarks considered, accurately recovering the underlying community structure.

\subsection{Resolution Limit Analysis on a Benchmark Network}
\label{ResolutionLimitBenchmark}

The resolution limit (RL) problem arises when an objective function incorrectly merges
well-defined communities~\cite{Newman2010book}. A resolution-limit-free method should separate such
groups while avoiding the artificial fragmentation of strongly connected ones.
\begin{figure}
\centering  
\includegraphics[width = 8 cm]{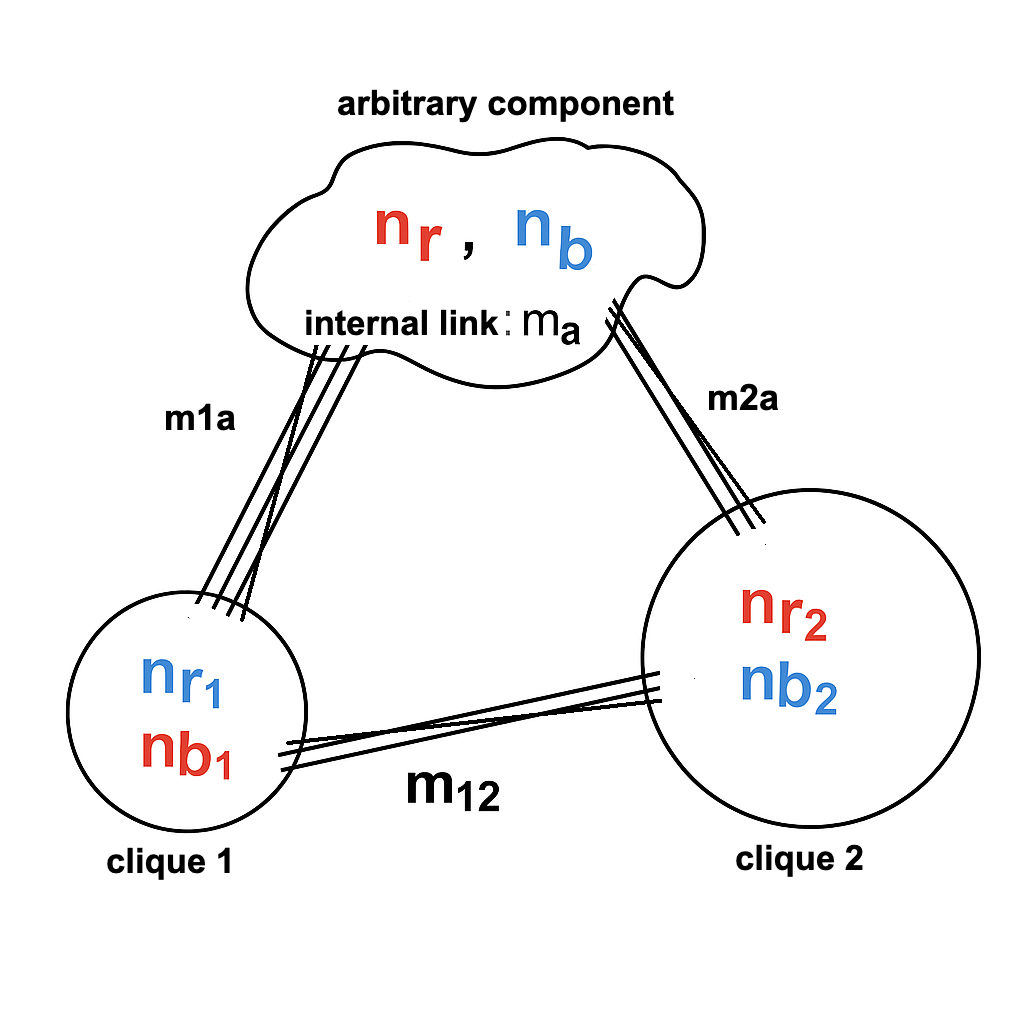}
\caption{\label{fig:BenchmarkNetwork}Benchmark network designed to study the resolution limit problem. It consists of two fully connected bipartite cliques of sizes \(n_1\) and \(n_2\), and an arbitrary external component containing \(n_a\) nodes and total internal weight \(m_a\). The cliques are weakly connected to this external component via \(m_{1a}\) and \(m_{2a}\) weighted links, respectively. Additionally, the cliques are connected to each other via \(m_{12}\) links. If the network is weighted, then \(m_a\), \(m_{1a}\), \(m_{2a}\), and \(m_{12}\) represent the sum of link weights in their respective regions.}
\end{figure}

We analyze this issue using the benchmark network shown in Fig.~\ref{fig:BenchmarkNetwork}, which
is a bipartite extension of benchmark constructions introduced in~\cite{Guo_2023,Lancichinetti_2011}.
The network consists of two fully connected bipartite cliques and an external arbitrary component
to which the cliques are weakly connected.

Let clique 1 contain \(n_{r1}\) red nodes and \(n_{b1}\) blue nodes, and clique 2 contain
\(n_{r2}\) red nodes and \(n_{b2}\) blue nodes. Without loss of generality, we assume
\(n_{r2} n_{b2} \geq n_{r1} n_{b1}\). Let \(m_{12}\) denote the total weight of links between the two
cliques, \(m_{1a}\) and \(m_{2a}\) the total weights of links connecting each clique to the external
component, and \(m_a\) the total internal weight of the external component, which contains \(n_a\)
nodes. We assume the cliques are weakly connected to the external component, such that
\(m_{1a} \ll n_{r1} n_{b1}\) and \(m_{2a} \ll n_{r2} n_{b2}\).

To characterize the network structure, we introduce the normalized inter-clique link density
\[
d = \frac{m_{12}}{\sqrt{n_{r1} n_{b1} n_{r2} n_{b2}}},
\]
the external influence parameter
\[
t = \frac{m_a}{\sqrt{n_{r1} n_{b1} n_{r2} n_{b2}}},
\]
the relative size ratio
\[
p = \frac{\sqrt{n_{r1} n_{b1}}}{\sqrt{n_{r2} n_{b2}}},
\]
and the clique size parameter
\[
r = p + \frac{1}{p}
  = \frac{n_{r1} n_{b1} + n_{r2} n_{b2}}
         {\sqrt{n_{r1} n_{b1} n_{r2} n_{b2}}}.
\]

The parameter \(p \in (0,1]\) quantifies the relative sizes of the two cliques, while
\(t \in [0,\infty)\) measures the strength of the external component. For a given objective
function, there exists a critical inter-clique link weight \(m_{\mathrm{exp}}\) that determines
whether the two cliques are merged or remain separated. Specifically, the cliques should be merged
when \(m_{12} \geq m_{\mathrm{exp}}\) and separated otherwise. We define the normalized expected
critical density
\[
\delta_{\mathrm{exp}}
= \frac{m_{\mathrm{exp}}}{\sqrt{n_{r1} n_{b1} n_{r2} n_{b2}}},
\]
which enables resolution-independent comparison across networks of different sizes.

For a fixed set of parameters, the optimization outcome partitions the parameter space into a
Merged (M) phase and a Split (S) phase, separated by a critical value \(\delta\). An objective
function is said to be resolution-limit-free with respect to the expected density if
\(\delta \geq \delta_{\mathrm{exp}}\) for all parameter configurations of the benchmark network.

\section{\label{sec:Results}Results}
\subsection{\label{sec:Results_A}Benchmark test}
We analytically examine the resolution-limit (RL) behavior of the benchmark network shown in
Figure~\ref{fig:BenchmarkNetwork} using \( Q_{bg} \) as the objective function. This analysis
allows us to determine whether \( Q_{bg} \) favors merging or splitting the two bipartite
cliques under different network configurations, and to compare its behavior with that of
other modularity-based objective functions.

Whether \( Q_{bg} \) favors merging or splitting the cliques is determined by the sign of the
quantity
\begin{equation}
    \Delta Q_{bg} = Q^{\text{merge}}_{bg} - Q^{\text{split}}_{bg},
    \label{eqtn:barber}
\end{equation}
where \( Q^{\text{merge}}_{bg} \) and \( Q^{\text{split}}_{bg} \) are the values of \( Q_{bg} \)
when the cliques are merged and split, respectively. If \( \Delta Q_{bg} < 0 \), the split
partition yields a higher objective value and the cliques remain separate; if
\( \Delta Q_{bg} > 0 \), the merged partition is preferred and the cliques are grouped into a
single community. This criterion provides a clear way to assess how \( Q_{bg} \) handles
community resolution across different benchmark configurations.

Then for $Q_{bg}$,
\small{
\begin{equation}
\Delta Q_{bg} \sim \left( \frac{d + r}{r + 2} \right)^{\chi} \left( d + r - \frac{(d + r)^2}{r + t + d} \right) - 
\left( r - \frac{rd + r^2 - 2}{r + t + d} \right)
\label{eq:delQ_bg}
\end{equation}
}
where \( r = p + \frac{1}{p} \). Equation~\ref{eq:delQ_bg} determines whether, for a given value
of the control parameter \( \chi \), the system lies in the Merged (M) or Split (S) phase as a
function of the parameters \( (p, d, t) \). The threshold link density, denoted
\( \delta_{Q_{bg}} \), corresponds to the critical value of \( d \) at which the transition
between the M and S phases occurs, that is, the point where \( \Delta Q_{bg} = 0 \).

When the external influence parameter \( t \) becomes large, a situation commonly encountered
in practical applications, the asymptotic form of \( \delta_{Q_{bg}} \) for a fixed \( \chi \)
is given by
\begin{equation}
    \lim_{t \to \infty} \delta_{Q_{bg}} \to r \left( \left(1 + \frac{2}{r} \right)^{\frac{\chi}{\chi + 1}} - 1 \right)
    \label{eqtn:del_Qbg}
\end{equation}

\begin{figure}
\centering  
\includegraphics[width = 8.6 cm]{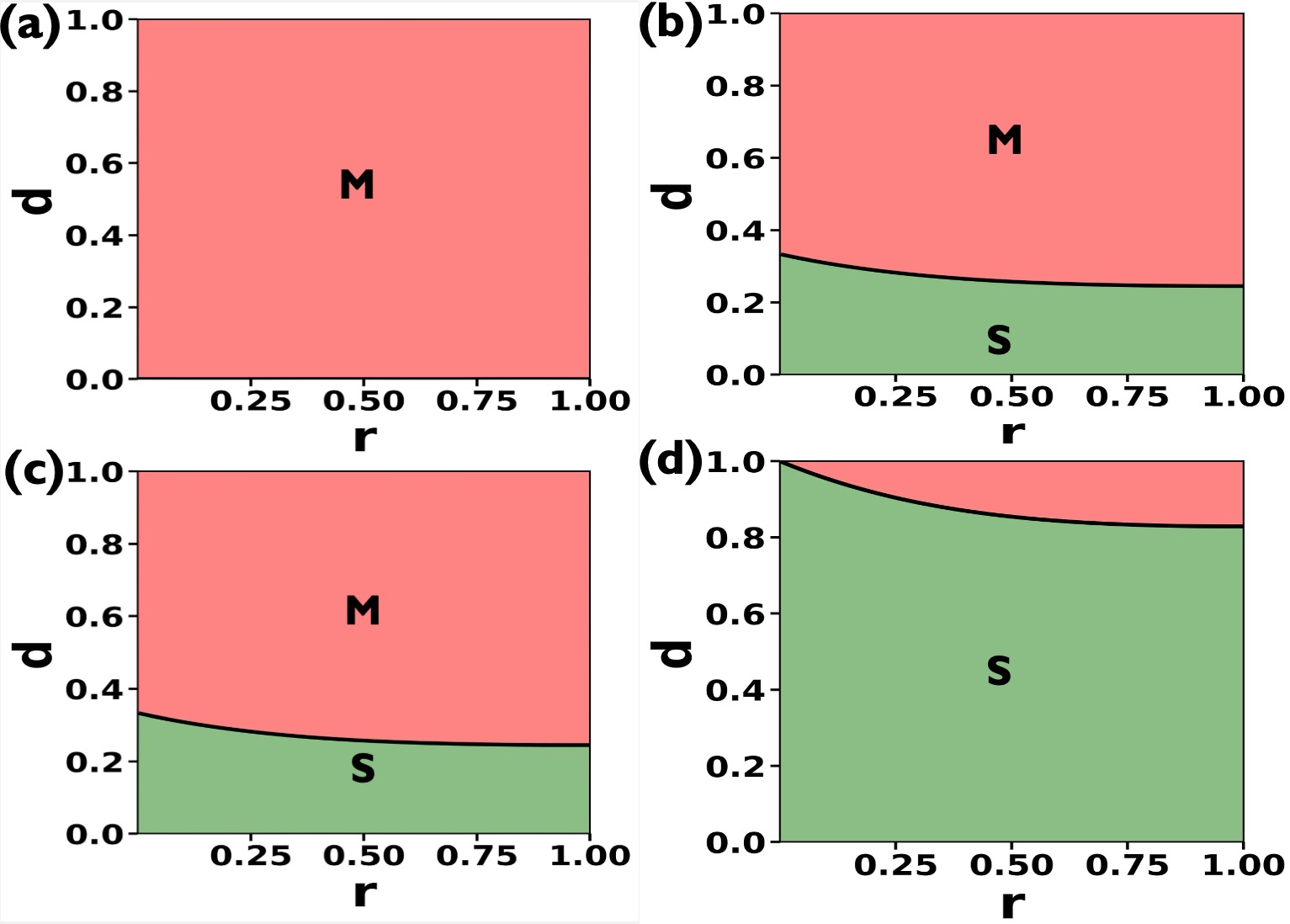}
\caption{\label{fig:Picture2}Phase diagram of clique splitting using $Q_{bg}$ at large external influence as \( \chi \) varies. The regions where the M phase occurs are shown in orange, and the regions where the S phase occurs are shown in blue. Results are presented for different values of the control parameter \( \chi \): (a) \( \chi = 0 \), (b) \( \chi = 0.2 \), (c) \( \chi = 0.5 \), and (d) \( \chi = 1.0 \). The external influence parameter is set to \( t = 10^6 \).
}
\label{fig:Picture2}
\end{figure}

\begin{figure}
\centering  
\includegraphics[width = 8.5 cm]{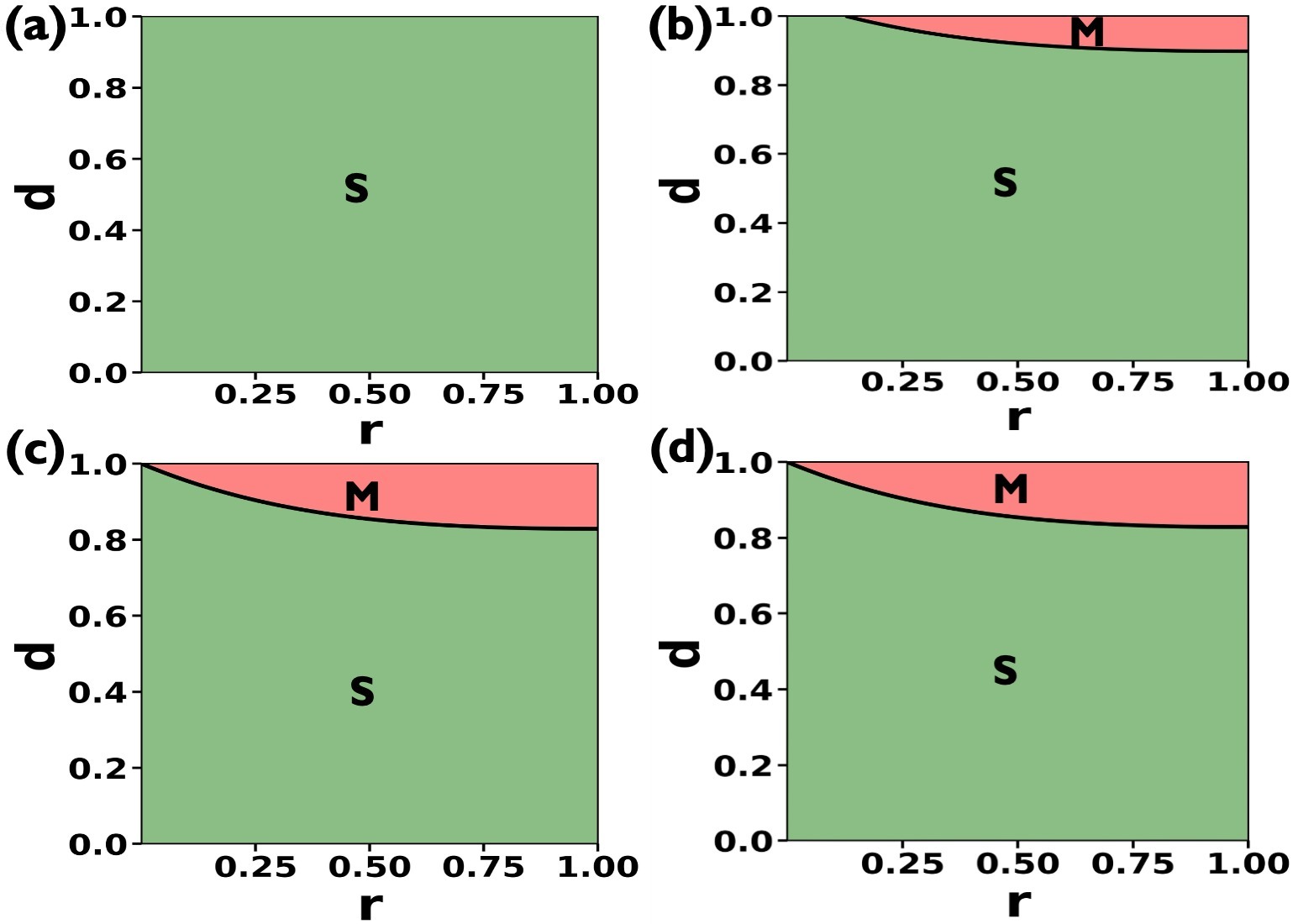}
\caption{\label{fig:Picture3}Phase diagram of clique splitting using $Q_{bg}$ at fixed \( \chi = 1 \), as the external influence parameter \( t \) is varied. The regions in the \((p, d)\) space where the Merged (M) phase occurs are indicated in orange, while the Split (S) phase regions are shown in blue. Results are presented for different values of \( t \): (a) \( t = 0 \), (b) \( t = 10 \), (c) \( t = 10^4 \), and (d) \( t = 10^6 \).
}
\end{figure}

The limiting value of \( \delta_{Q_{bg}} \) increases from 0 when \( \chi = 0 \) and approaches 1
as \( \chi \to \infty \), independent of the value of \( p \). For intermediate values of
\( \chi \), the dependence on \( p \) is weak, with slightly higher critical values observed at
smaller \( p \), as illustrated in Figure~\ref{fig:Picture2}. This figure shows the phase
diagram in the \( (p, d) \) plane for several values of \( \chi \) under conditions of large
external influence. When \( \chi = 0 \), the cliques remain merged across the parameter space
shown in Figure~\ref{fig:Picture2}(a). For \( \chi > 0 \), the cliques become separated at lower
values of \( d \), indicating that they are successfully resolved. As \( \chi \) increases, the
critical density \( \delta_{Q_{bg}} \) increases and approaches 1, as seen in
Figures~\ref{fig:Picture2}(b)--(d). This implies that for sufficiently large \( \chi \), the
cliques are resolved throughout the parameter space.

The effect of varying the external influence parameter \( t \) at fixed \( \chi \) is shown in
Figure~\ref{fig:Picture3}. When \( t = 0 \), as shown in
Figure~\ref{fig:Picture3}(a), the external component is absent and the S phase occupies the
entire \((p, d)\) space. In this case, \( \delta_{Q_{bg}} = 1 \) for all values of \( p \),
implying that the cliques remain separated unless they are fully connected. When \( t > 0 \),
the presence of the external component promotes merging at sufficiently large values of \( d \),
reflecting a loss of resolution. As \( t \) increases, shown in
Figures~\ref{fig:Picture3}(b)--(d), the M phase expands and \( \delta_{Q_{bg}} \) decreases,
eventually approaching the limiting value given by Equation~\ref{eqtn:del_Qbg}.

These results demonstrate that adjusting the control exponent \( \chi \) provides a direct way
to tune \( \delta_{Q_{bg}} \) over a wide range. This tunability becomes broader as \( t \)
increases, and in the limit of large \( t \), \( \delta_{Q_{bg}} \) spans the full interval from
0 to 1. This flexibility is advantageous in applications, as it allows \( \chi \) to be chosen
such that \( \delta_{Q_{bg}} \) matches the expected critical resolution density
\( \delta_{\text{exp}} \).

As \( \chi \) increases, the number of detected communities generally increases, while the
results stabilize over a range of \( \chi \) values (see the example in
Section~\ref{subsec:artificial-hierarchical-bipartite}). Increasing \( \chi \) typically leads
to the detection of smaller communities. The optimal choice of \( \chi \) depends on the
specific problem under consideration. When prior information about the community structure is
available, it can guide the selection of \( \chi \). In the absence of such information,
\( \chi = 1 \) provides a reasonable default choice. A key advantage of the \( Q_{bg} \) metric
is that even for extreme values of \( t \) and \( r \), it maintains a positive lower bound
\( \delta_{Q_{bg}} \), which can be adjusted through the choice of \( \chi \).

\subsection{\label{Applications}Applications}
\subsubsection{\label{sec:Results_B}Artificial hierarchical bipartite network}
\label{subsec:artificial-hierarchical-bipartite}

To examine how the proposed generalized bipartite modularity density \(Q_{bg}\) captures
multiscale community structure, we construct an artificial hierarchical bipartite network
designed to exhibit clear organization across multiple resolution levels. Inspired by
hierarchical network models~\cite{Arenas_2008,Guo_2023}, the construction consists of four
nested levels, each corresponding to a distinct structural scale.

\begin{figure}
\centering
\includegraphics[width = 8.5 cm]{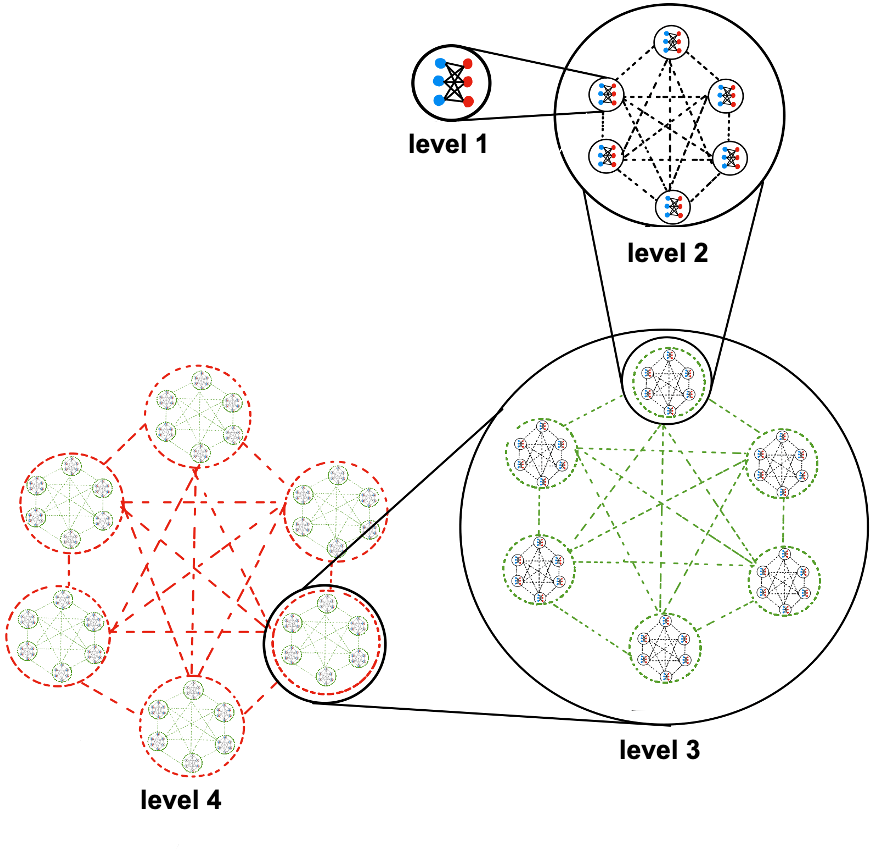}
\caption{\label{fig: ArtificialBip}
Artificial hierarchical bipartite network.
Level~1: A bipartite clique of three blue and three red nodes.
Level~2: A clique of five Level~1 cliques.
Level~3: A clique of five Level~2 cliques.
Level~4: A clique of five Level~3 cliques.
}
\label{Artificialnetwork}
\end{figure}

At Level~1, the basic building block is a bipartite clique in which all red nodes are fully
connected to all blue nodes, with each link assigned a weight \(\alpha_1\). At Level~2,
these Level~1 cliques are treated as generalized nodes and assembled into a larger clique,
with full interconnections weighted by \(\alpha_2\). This procedure is iterated such that
the Level~\(k\) structure is formed by treating each Level~\(k-1\) network as a generalized
node and arranging them into a clique of size five, with interconnections weighted by
\(\alpha_k\). To preserve a clear hierarchical organization, the link weights are chosen to
satisfy \(\alpha_1 > \alpha_2 > \cdots > \alpha_k\), ensuring that connections weaken
systematically at higher levels of the hierarchy. A schematic illustration of this
construction is shown in Fig.~\ref{fig: ArtificialBip}.

\begin{figure}
\centering
\includegraphics[width = 8.5 cm]{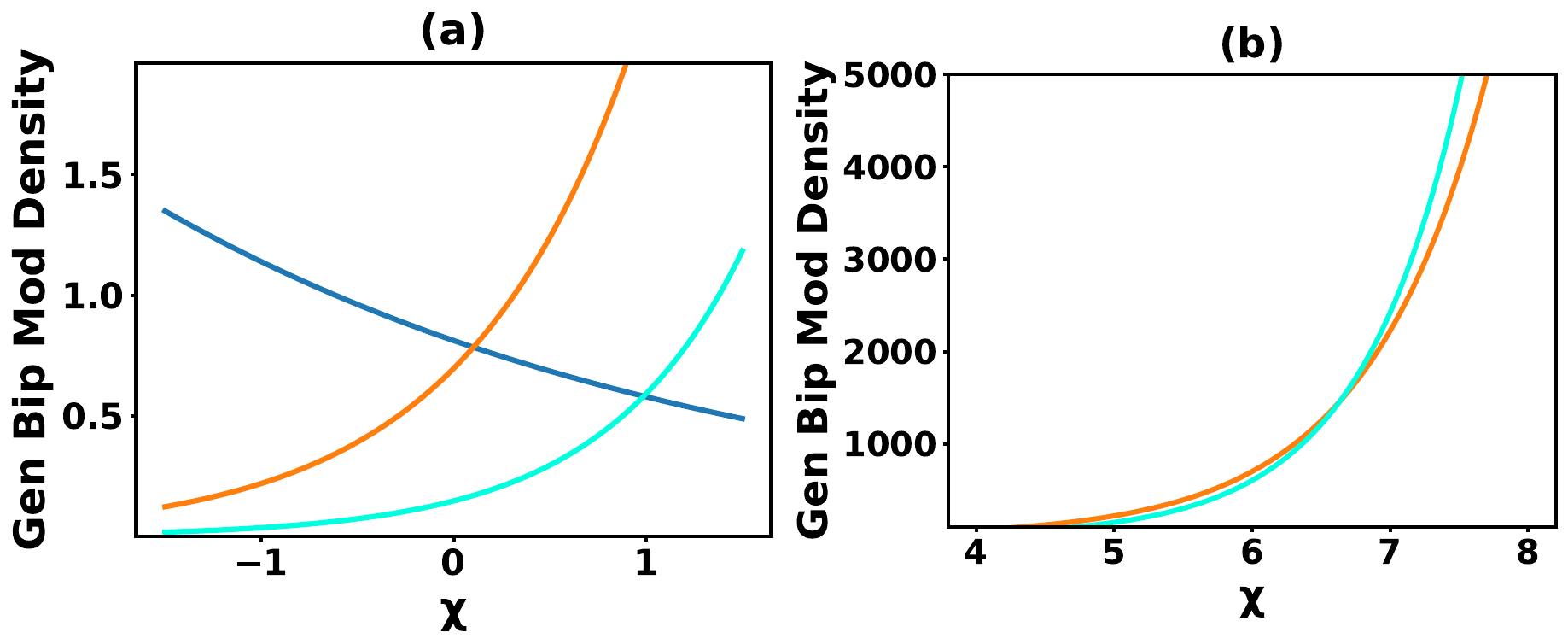}
\caption{\label{fig: communities}
Multiresolution community detection in the artificial hierarchical bipartite network.
The generalized bipartite modularity density $Q_{bg}$ is plotted as a function of the
resolution parameter $\chi$, with curves corresponding to partitions into $6$ cliques
(blue), $36$ cliques (orange), and $216$ cliques (green).
\textbf{(a)} Low-resolution regime, $-1.5 \le \chi \le 1.5$.
In this regime, for $\chi < 0.15$, the blue curve attains the highest value of $Q_{bg}$,
indicating that the partition into six Level-3 cliques is optimal.
\textbf{(b)} Higher-resolution regime.
For $0.15 \le \chi < 6.5$, the orange curve dominates, corresponding to the emergence
of $36$ Level-2 cliques.
For $\chi \ge 6.5$, the green curve yields the maximum $Q_{bg}$, resolving the finest-scale
structure of $216$ Level-1 cliques.
Together, panels (a) and (b) illustrate how varying $\chi$ reveals community structure
from the largest to the smallest scales in the hierarchical bipartite network.
}
\end{figure}

We apply the \(Q_{bg}\) metric to a Level~4 hierarchical network constructed using
\(\alpha_k = 5 - k\). The resulting network contains 216 Level~1 cliques, comprising a total
of 648 red nodes and 648 blue nodes. By maximizing \(Q_{bg}\), we demonstrate that the planted
community structure at each hierarchical level can be successfully recovered. The
communities identified at different resolutions are shown in
Fig.~\ref{fig: communities}, where distinct partitions emerge as the resolution parameter
\(\chi\) is varied.

\begin{figure}
\centering
\includegraphics[width = 8cm]{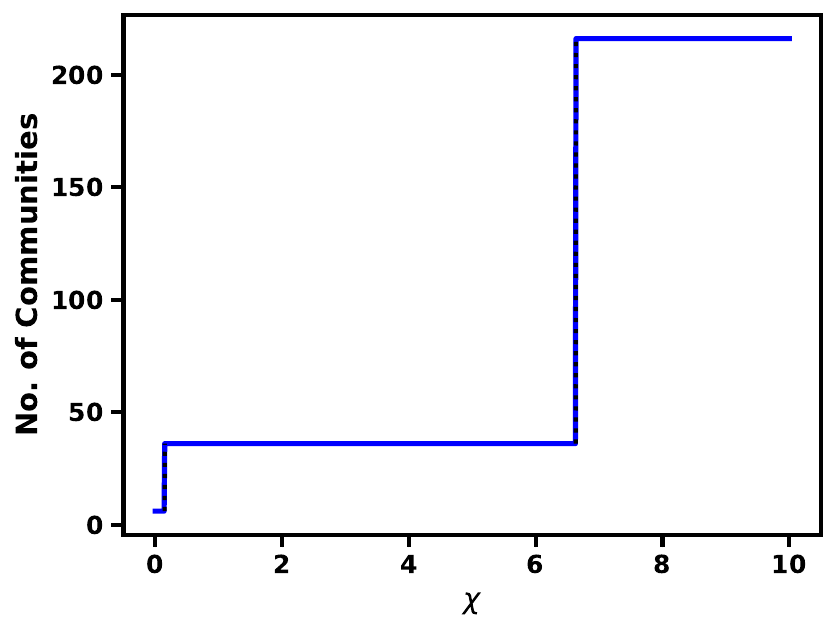
}
\caption{\label{fig:community_chi}
The number of communities detected using $Q_{bg}$ varies with different values of $\chi$.
As $\chi$ increases, communities are progressively uncovered from the largest to the
smallest.
}
\end{figure}

Specifically, the six Level~3 cliques are detected for \(\chi < 0.15\), the 36 Level~2 cliques
emerge for \(0.15 \le \chi < 6.5\), and the 216 Level~1 cliques are resolved for \(\chi \ge 6.5\).
These transitions define three distinct regimes, each corresponding to a different level
of the hierarchical construction. The dependence of the number of detected communities on
the resolution parameter \(\chi\) is summarized in Fig.~\ref{fig:community_chi}, which
illustrates how finer structures are progressively uncovered as \(\chi\) increases.

Due to the inherently multiscale nature of the hierarchical network, there is no single
optimal choice of \(\chi\). Instead, the resolution parameter should be selected based on
the scale of interest and any available prior knowledge about the system. This example
demonstrates the flexibility of \(Q_{bg}\) in uncovering meaningful community structure
across multiple hierarchical levels.
\subsection{Real bipartite networks}
\label{Real bipartite networks}

\subsubsection{Southern women bipartite network}

\begin{figure}
\centering
\includegraphics[width = 8.5cm]{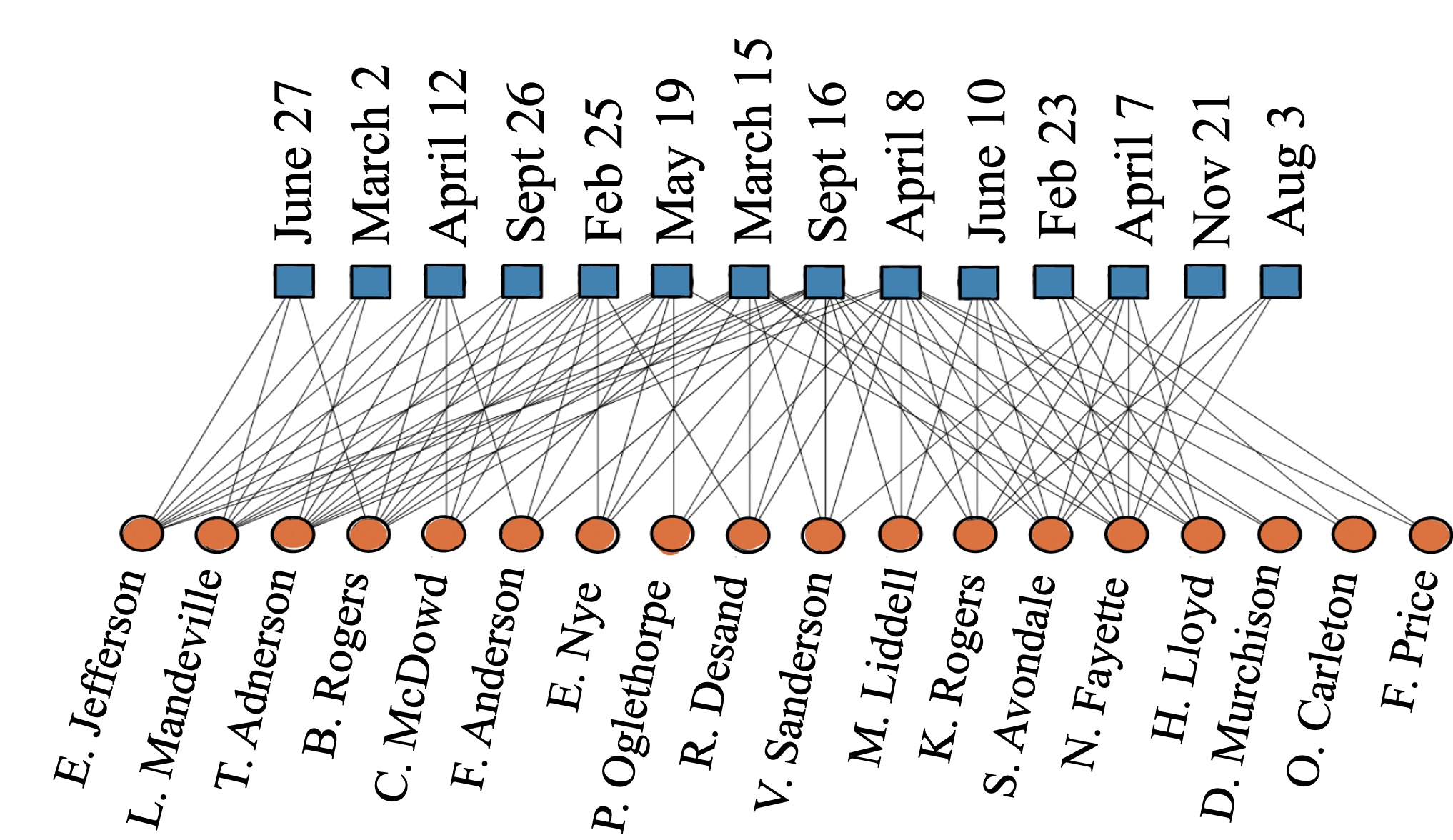}
\caption{\label{fig:SouthernWomen}
Bipartite representation of the Southern Women network. Circles represent women and squares represent social events; an edge connects a woman to an event she attended. The network encodes attendance patterns of 18 women at 14 events in a Southern U.S.\ town during the 1930s~\cite{davis1941deep}.}
\end{figure}

We apply the bipartite modularity density metric \( Q_{bg} \) to the well-known Southern Women bipartite network, which records attendance patterns of 18 women at 14 social events in a Southern U.S.\ town during the 1930s~\cite{davis1941deep}. Figure~\ref{fig:SouthernWomen} shows the bipartite representation of this network, where nodes correspond to women and social events, and edges indicate event attendance. In this network, edges represent participation of women in events. Because many women attended similar subsets of events, an effective community detection method should be able to identify groups of women with comparable attendance behavior, as well as the events that define those groups.

\begin{figure}
\centering
\includegraphics[width = 9.5cm]{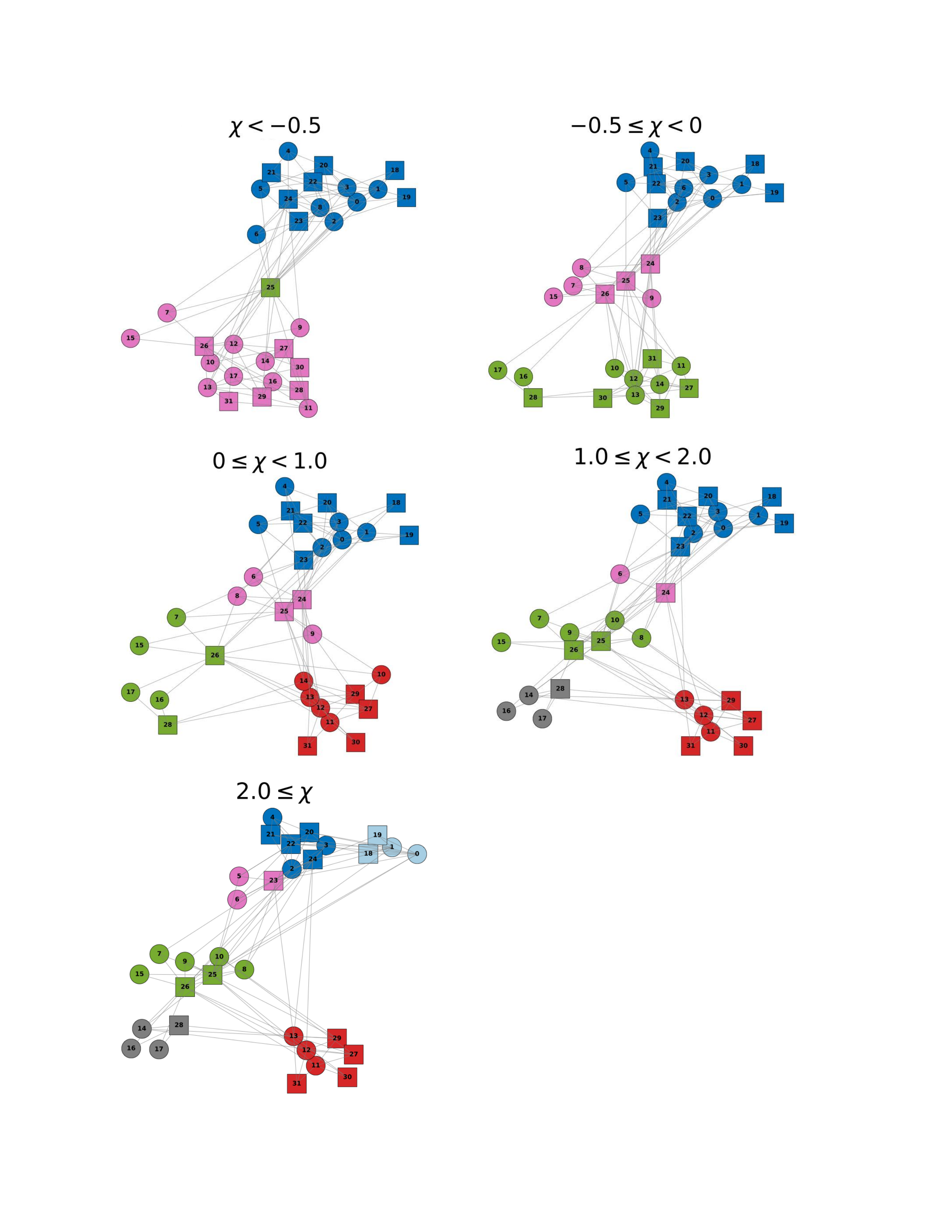}
\caption{\label{fig:FinalMerge}
Hierarchical clustering of the Southern Women event network with increasing $\chi$.}
\end{figure}

We examine how the detected community structure evolves as the resolution parameter $\chi$ varies from $-1.0$ to $2.0$, as shown in Fig.~\ref{fig:FinalMerge}. An animated visualization illustrating the transitions of nodes between partitions across resolutions is available at
\href{https://youtube.com/shorts/ZqRcboO0Cb8?si=u5I1tuaLcZtNRpTO}{this link}.

For the lowest resolution range, $-1.0 \leq \chi < -0.5$, the women are broadly divided into two communities. This coarse partition closely corresponds to the primary division identified in the original ethnographic study~\cite{davis1941deep}, which described two major social groups within the town.

As the resolution increases to $-0.5 \leq \chi < 0$, the network separates into three communities. Two of these correspond to the main groups of socially active women, while the third (shown in pink) consists of women who attended relatively few events. According to the ground-truth analysis~\cite{davis1941deep}, these women occupy a peripheral position in the social structure: they participate infrequently, have weaker social ties, and do not strongly affiliate with either of the main groups. The emergence of this peripheral cluster demonstrates that $Q_{bg}$ is sensitive to differences in participation intensity at intermediate resolutions.

For $0 \leq \chi < 1.0$, the detected partition exactly matches the canonical community structure reported by Barber~\cite{Barber_2007}. This agreement provides strong validation of our method, showing that $Q_{bg}$ recovers well-established bipartite community divisions without requiring projection to a unipartite network.

As $\chi$ is increased further, finer-grained structure becomes visible. At higher resolutions, the network splits into five communities, including a newly resolved cluster primarily composed of core members of the second major social group identified in the original study~\cite{davis1941deep}. At $\chi = 2.0$, the algorithm identifies six smaller communities, including a new group formed by core members of the first major subgroup. These results indicate that increasing the resolution parameter reveals progressively finer subdivisions within the broader social organization.

Overall, the patterns uncovered by our method are consistent with those reported in the original ethnographic analysis~\cite{davis1941deep} and subsequent studies~\cite{Barber_2007}. However, prior work provided only partial and resolution-specific views of the community structure. In contrast, our approach offers a unified hierarchical perspective across the full range of $\chi$, revealing how communities emerge, split, and refine as resolution changes. To the best of our knowledge, this represents the first complete multiscale characterization of the Southern Women bipartite network.

It is important to emphasize that our method does not cluster women alone; rather, it identifies \emph{biclusters}. Each detected community consists of a group of women together with the specific events that connect them. In this way, the algorithm simultaneously groups women with similar attendance patterns and the events that define those patterns. By preserving the full bipartite structure, this biclustering approach avoids the information loss inherent in unipartite projections and provides a more faithful representation of the joint social--event organization.

These results demonstrate that the proposed metric $Q_{bg}$ is highly effective for uncovering hierarchical community structure in bipartite networks and holds strong promise for applications to other real-world bipartite systems.

\subsubsection{Asthma patient bipartite network}
\label{Asthma_network}

We apply our bipartite community detection method based on the modularity density function $Q_{bg}$ to an asthma-related bipartite network consisting of 83 patients and 18 cytokines, where edges represent significant cytokine activity levels measured in patients~\cite{Bhavnani2012}. The original biological study identified three major patient groups: (i) a group associated with IL-4 and Eotaxin, (ii) a group enriched in IL-5, IFN-$\gamma$, IL-17, MIP-$1\alpha$, MIP-$1\beta$, and MIG, and (iii) a group characterized by generally low cytokine expression.

\begin{figure}
\centering
\includegraphics[width = 8.5cm]{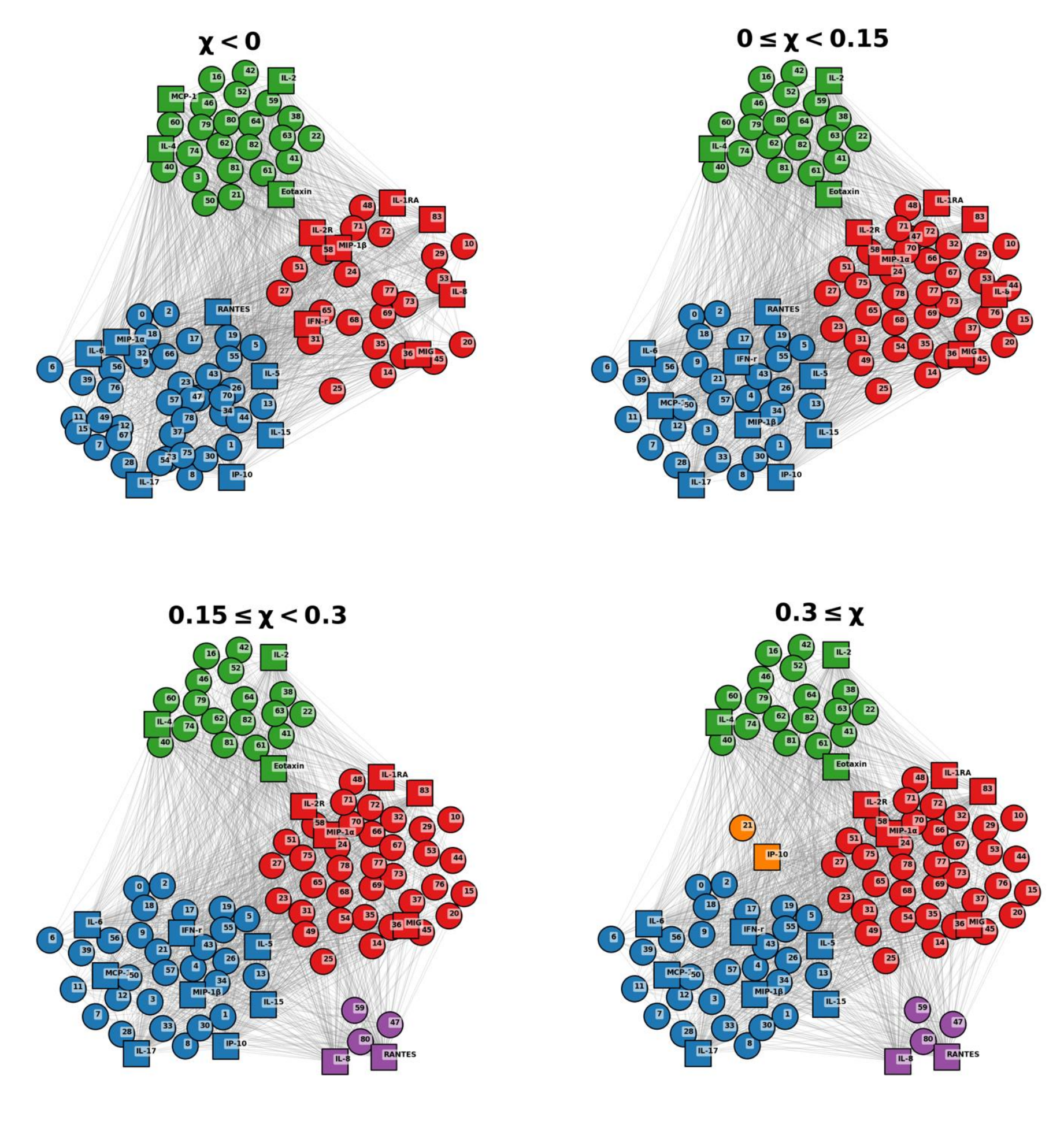}
\caption{\label{fig:Asthma}
Hierarchical clustering of the asthma patient--cytokine bipartite network with increasing $\chi$.}
\end{figure}

Figure~\ref{fig:Asthma} shows the detected community structure as the resolution parameter $\chi$ varies from $-2.0$ to $0.3$. An animated visualization illustrating how patients and cytokines transition between communities across resolutions is available at
\href{https://youtube.com/shorts/_3uuvS56v74}{this link}.

For low resolution values, $-2.0 \leq \chi < 0$, our method identifies three broad clusters, consistent with the coarse partition reported in the original study~\cite{Bhavnani2012}. As $\chi$ increases, the detected communities align more closely with biologically meaningful cytokine groupings. In particular, we observe a strong and persistent association between IL-4 and Eotaxin, reflecting a T-helper-2 ($T_H2$) immune response in asthma, where IL-4 induces Eotaxin production and promotes eosinophil recruitment~\cite{Bhavnani2012}.

Notably, IL-2 is consistently clustered with IL-4 and Eotaxin across a wide range of $\chi$ values. Although not emphasized in the original study~\cite{Bhavnani2012}, this association is supported by the underlying data: patients with elevated IL-4 and Eotaxin levels also tend to exhibit increased IL-2 expression, suggesting coordinated immune activation within this cluster.

At $\chi = 0$, we observe a stable cluster containing IL-5, IFN-$\gamma$, IL-17, and MIP-$1\beta$, in agreement with the original analysis~\cite{Bhavnani2012}. This cytokine group is regulated by the NF-$\kappa$B pathway, which is associated with enhanced pro-inflammatory responses and more severe asthma phenotypes. In contrast, MIG and MIP-$1\alpha$, which were previously grouped with this cluster, instead form a separate community in our analysis. These cytokines merge with IL-15, TNF-$\alpha$, and IL-1Ra, forming a distinct pro-inflammatory subgroup, suggesting additional heterogeneity in cytokine co-expression patterns.

As $\chi$ increases further ($\chi \geq 0.15$), new structures emerge that were not explicitly reported in the original study. In particular, RANTES and IL-8 form a distinct cluster, accompanied by a subgroup of patients exhibiting elevated expression of these cytokines. At higher resolutions ($\chi > 0.3$), IP-10 separates into its own cluster, driven primarily by a single patient with exceptionally high IP-10 expression.

In general, our results provide a multiresolution view of patient--cytokine interactions that recovers the major groupings reported in~\cite{Bhavnani2012}, while also revealing additional and potentially meaningful immune signaling patterns. The persistent clustering of IL-2 with IL-4 and Eotaxin, the emergence of a RANTES--IL-8 cluster, and the isolation of IP-10 at higher resolutions highlight the ability of $Q_{bg}$ to uncover the hierarchical structure and biological heterogeneity. Importantly, the method performs true biclustering, simultaneously grouping patients with the cytokines they express most strongly, providing a more complete representation of asthma immune phenotypes.


\appendix

\section{Resolution Limit in Bipartite Clique Ring}
\label{sec:appendix_resolution}

In Ref.~\cite{ABC}, the authors derived a criterion describing when bipartite modularity fails to resolve individual communities in a bipartite clique ring. Upon re-examination, however, we find that their derivation is incorrect. Here, we present the corrected condition determining when a bipartite $k$-clique ring exhibits the resolution limit.

Starting from Eq.~\ref{eq:eqtn2}, we obtain

{\small
\begin{align}
n\left(\frac{k^2}{n + nk^2} - \frac{(k^2 + 1)^2}{(n + nk^2)^2}\right)
&> \frac{n}{2} \left( \frac{2k^2 + 1}{n + nk^2} - \frac{4(k^2 + 1)^2}{(n + nk^2)^2} \right) \notag \\
\Rightarrow\quad
\frac{2k^2}{2(n + nk^2)} - \frac{2k^2 + 1}{2(n + nk^2)} 
&> -\frac{(k^2 + 1)^2}{(n + nk^2)^2} \notag \\
\Rightarrow\quad
\frac{1}{2} 
&< \frac{(k^2 + 1)^2}{(n + nk^2)} \notag \\
\Rightarrow\quad
k &> \sqrt{\frac{n - 2}{2}} 
\label{eq:reslimit_condition}
\end{align}
}

Thus, bipartite modularity remains resolution-limit-free only when

\begin{equation}
k > \sqrt{\frac{n - 2}{2}} .
\label{eq:reslimit_threshold}
\end{equation}

We illustrate this condition with two examples:

\begin{enumerate}
    \item For \( n = 4 \) and \( k = 2 \),
    \[
    Q_b(M) = 0.4 \quad \text{and} \quad Q_b(S) = 0.55,
    \]
    so $Q_b(S) > Q_b(M)$ and the network is resolution-limit-free.

    \item For \( n = 12 \) and \( k = 2 \),
    \[
    Q_b(M) = 0.733 \quad \text{and} \quad Q_b(S) = 0.716,
    \]
    giving $Q_b(S) < Q_b(M)$ and demonstrating the resolution limit.
\end{enumerate}

\section{Phase diagrams illustrating clique splitting behavior under modularity $Q_b$ in Benchmark Network}
\label{sec:appendix_phase}

The respective values of $Q_b$ resulting from splitting or merging the two cliques in the benchmark network are

\begin{align}
    Q_{b}^{\text{split}} &= Q_{b1} + Q_{b2} + Q_{b(\text{ex})}  \label{eq:qb_split}\\
    Q_{b}^{\text{merge}} &= Q_{b(1+2)} + Q_{b(\text{ex})}        \label{eq:qb_merge}
\end{align}

Here, $Q_{b1}$ and $Q_{b2}$ denote the modularity contributions of cliques 1 and 2 when they are treated as separate communities, whereas $Q_{b(1+2)}$ corresponds to the merged configuration. The term $Q_{b(\text{ex})}$ represents the contribution from the external component, which remains unchanged in both cases.

Thus, the modularity difference between the merged and split configurations is
\begin{equation}
    \Delta Q_b = Q^{\text{merge}}_b - Q^{\text{split}}_b
               = Q_{b(1+2)} - Q_{b1} - Q_{b2},
    \label{eq:qb_delta_def}
\end{equation}

and we obtain
\begingroup
\scriptsize
\begin{align}
\Delta Q_b
&= \frac{1}{F} \left[
      \left( m_{(1+2)} - \frac{q_{(1+2)} d_{(1+2)}}{F} \right)
      - \left( m_1 - \frac{q_1 d_1}{F}
             + m_2 - \frac{q_2 d_2}{F} \right)
   \right]
   \label{eq:qb_delta_general} \\
&= \frac{1}{F}
   \left(
      m_{12}
      - \frac{
          2 n_{r_1} n_{b_1} \, n_{r_2} n_{b_2}
          + (n_{r_1} n_{b_1} + n_{r_2} n_{b_2}) m_{12}
          + m_{12}^2
        }{
          n_{r_1} n_{b_1}
          + n_{r_2} n_{b_2}
          + m_{12}
          + m_a
        }
   \right)
   \label{eq:qb_delta_expanded}
\end{align}
\endgroup

Equation~\ref{eq:qb_delta_expanded} can be rewritten in terms of the nondimensional
variables $(d,r,t)$ introduced in Section~\ref{ResolutionLimitBenchmark}, yielding
\begin{equation}
    \Delta Q_b \sim d - \frac{2 + r d + d^2}{r + d + t}.
    \label{eq:qb_delta_simplified}
\end{equation}

The phase behavior is determined by $(d,r,t)$, with
$0 \le d \le 1$, $r \ge 2$, and $t \ge 2$.
The phase boundary occurs when $\Delta Q_b = 0$, giving

\begin{equation}
\delta_Q{_b} = \frac{2}{t}
\label{eq:qb_delta_boundary}
\end{equation}

\begin{figure}[htbp]
\centering  
\includegraphics[width = 8.5 cm]{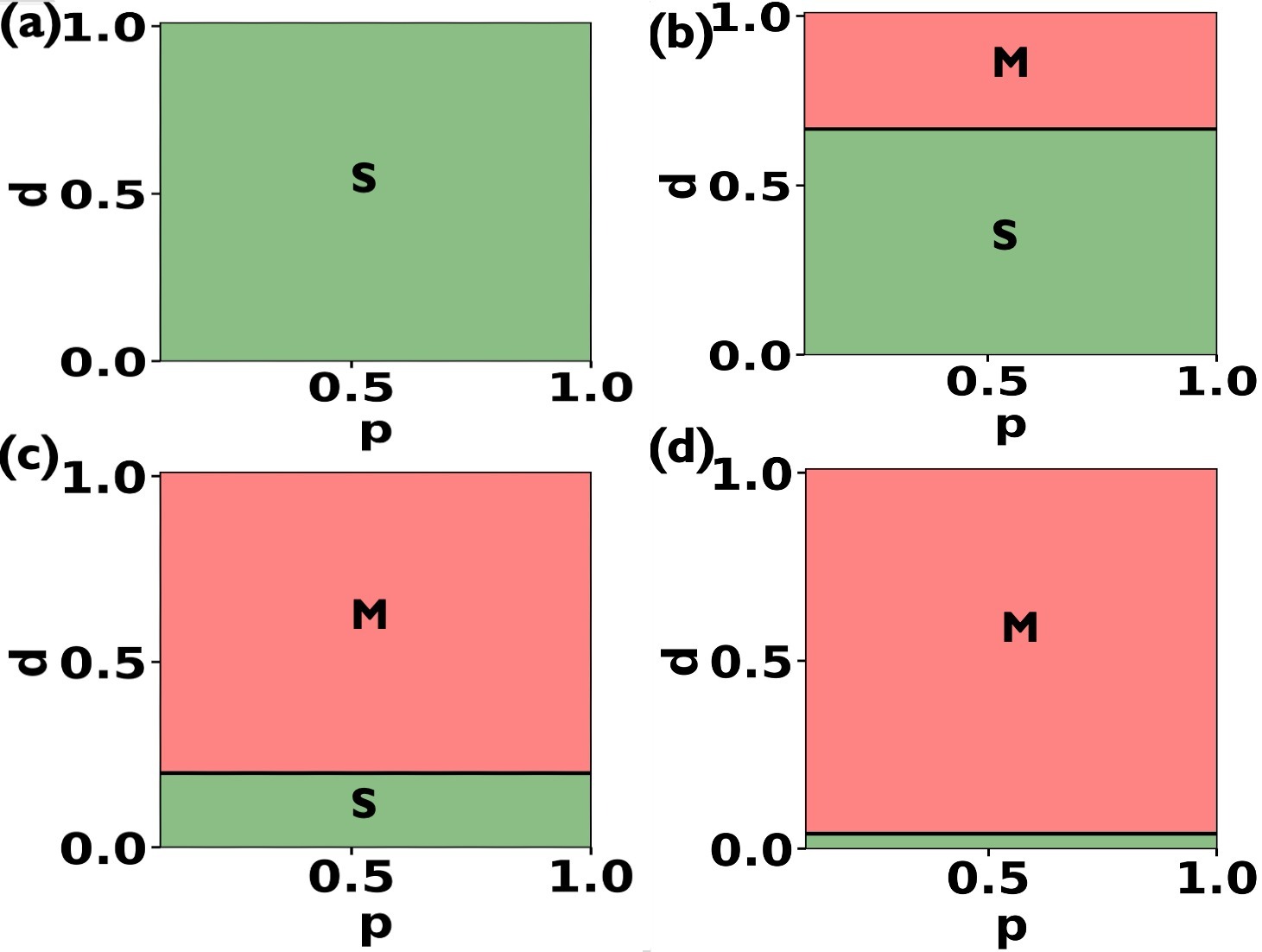}
\caption{\label{fig:appendix_qb_phase}Phase diagrams illustrating clique splitting behavior under modularity \( Q_b \) as the external influence parameter \( t \) is varied. Each panel shows the regions in the \( (p, d) \) space where the two cliques are either merged (M phase, orange) or split (S phase, blue), for \( t = 0, 3, 10, 50 \).}
\end{figure}
\nocite{*}

\bibliography{apssamp}

\end{document}